\documentclass[11pt,twoside]{article}

\usepackage{asp2006}
\usepackage{epsf}
\usepackage{psfig}
\usepackage{lscape}

\begin{document}
\title{Disk galaxies at z=2 in OWLS}   
\author{Julio F. Navarro$^{1}$, 
Laura V. Sales$^{2}$,
Joop Schaye$^{3}$,
Claudio Dalla Vecchia$^{3}$ and
Volker Springel$^{4}$}
\affil{
$^{1}$ Department of Physics and Astronomy, University of Victoria, Victoria, BC V8P 5C2,
Canada.\\
$^{2}$ Kapteyn Astronomical Institute, P.O. Box 800, Groningen, The Netherlands.\\
$^{3}$ Leiden Observatory, Leiden University, PO Box 9513, 2300 RA Leiden, The Netherlands\\
$^{4}$ Max Planck Institute for Astrophysics, Karl-Schwarzschild-Strasse 1, 85740 Garching, Germany\\
}    

\begin{abstract} We use the OWLS (OverWhelmingly Large Simulations)
set of cosmological Nbody/gasdynamical simulations to study the
properties of simulated galaxies at $z=2$. We focus on the effect of
supernova feedback from evolving stars on the baryonic mass and
angular momentum content of galaxies that assemble at the center of
$10^{11}$-$10^{12} \,h^{-1}M_\odot$ halos. Our main finding is that
the mass and angular momentum of such galaxies are strongly coupled,
in a way that is approximately independent of feedback: varying the
feedback implementation leads, in a given halo, to large variations in
galaxy mass but leaves the galaxy mass-angular momentum correlation
largely unaltered. In particular, the ratio between the angular
momentum of a galaxy and that of its surrounding halo ($j_d=J_{\rm
  gal}/J_{\rm vir}$) correlates closely with the galaxy mass
(expressed in units of the virial mass of the halo; $m_d=M_{\rm gal}/
M_{\rm vir}$). This correlation differs substantially from the
$m_d=j_d$ assumption commonly adopted in semianalytic models of galaxy
formation. We use these results to infer the sizes of disk galaxies at
$z=2$ expected in the LCDM scenario and to interpret recent
observations of extended disks at $z \sim 2$ by the SINS
collaboration.  \end{abstract}

\section{Introduction}  

It is widely thought that the properties of disk galaxies formed
hierarchically (as expected in the prevailing LCDM paradigm) are
largely determined by the properties of the dark matter halos in which
they form. In this scenario, the scaling laws linking various disk
galaxy properties, such as their mass, size, and rotation speed,
simply reflect analogous correlations between the mass, spin, and
potential depth of their surrounding halos (see, e.g., Navarro \&
Steinmetz 2000). Since the correlations linking various halo
properties have been studied extensively through N-body simulations,
the presumed association between disk and halo scaling laws enables a
host of predictions for the properties of disk galaxies and their
evolution with redshift. The general analytical framework has been
developed in some detail by Mo, Mao \& White (1998, hereafter MMW98),
who showed that a number of properties of the present-day disk
population may be reproduced with well-motivated choices for the small
number of free parameters in the model.

One important prediction of this modeling is that, at fixed rotation
speed, disk galaxies should become smaller with increasing
redshift. This prediction has come recently under scrutiny given the
results of the SINS collaboration, who report the discovery of a
population of extended disk galaxies at $z=2$ (Genzel et al. 2006,
F{\"o}rster Schreiber et al. 2006). In some cases, these galaxies have
rotation speeds comparable to $L_*$ disks today and are as extended as
their $z=0$ counterparts. This in apparent contradiction with the
predictions of the MMW98 modeling, unless these galaxies are embedded
in halos with unusually high angular momentum (Bouch{\'e} et al. 2007).

The interpretation of this disagreement is unclear, however, because
N-body/gasdynamical simulations of galaxy formation have shown that a
number of the assumptions made in the MMW98 model are not readily
verified in the simulations. In particular, the mass and angular
momentum of galaxies assembled hierarchically are not simple functions
of the mass and spin of their surrounding halos (Navarro \& Steinmetz
1997, 2000). The time of collapse; the importance of mergers; the
feedback from evolving stars; all of these effects can alter
dramatically the simple correspondence between galaxies and host halos
envisioned in semianalytic models (Okamoto et al. 2005; Robertson et
al. 2006; Governato et al 2007). As a result, it is unclear whether
the extended disks observed by the SINS collaboration present a
challenge for the LCDM paradigm or whether they may be accommodated by
reasonable tuning of the model free parameters.

We address these issues here using simulated galaxies selected from
the OverWhelmingly Large Simulations (OWLS) project. We use these
simulations to examine the relation between mass and angular momentum
(and its dependence on feedback) in $z=2$ galaxies assembled in the
LCDM scenario. We use these results to validate the MMW98 model and to
compare with the results from the SINS collaboration.
 
\section{The numerical simulations}  

The OverWhelmingly Large Simulation (OWLS) project is a collection of
roughly $50$ Nbody/gasdynamical cosmological simulations of
representative volumes in an LCDM universe. The various OWLS runs
explore systematically the effect on the simulated galaxy population
of varying numerical resolution, star formation laws, cooling
function, sub-grid physics, and feedback recipes.

We select for this study four different OWLS runs that differ from
each other only in the way supernova (SN) feedback is implemented.
Each run follows the evolution of $512^3$ dark mater and $512^3$ gas
particles in a $25 \, h^{-1} \rm Mpc$ box.  The mass per particle in
these runs is $\sim 6 \times 10^{6} \, h^{-1} \rm M_\odot$ for the
dark matter and six times lower for the baryons. Gas is allowed to
cool radiatively, and to turn into stars in a manner consistent with
an empirically calibrated Schmidt-Kennicutt law. Feedback energy from
evolving stars is incorporated assuming that a fixed fraction of the
available energy ($40\%$ of the total energy released by SN for a
Chabrier IMF) is channeled into bulk motions of the gas particles in
order to mimic feedback-driven ``winds'' outflowing from regions of
active star formation. In all four runs the same amount of energy is
invested in feedback; the main differences between runs are the
choices made for the wind velocity ($v_{\rm w}$) and mass loading ($\eta$)
factor. The parameter $\eta$ specifies the number of gas particles
among which the feedback kinetic energy is split, whereas $v_{\rm w}$
characterizes the outflow velocity of the wind particles.

\begin{figure}[!ht]
\plottwo{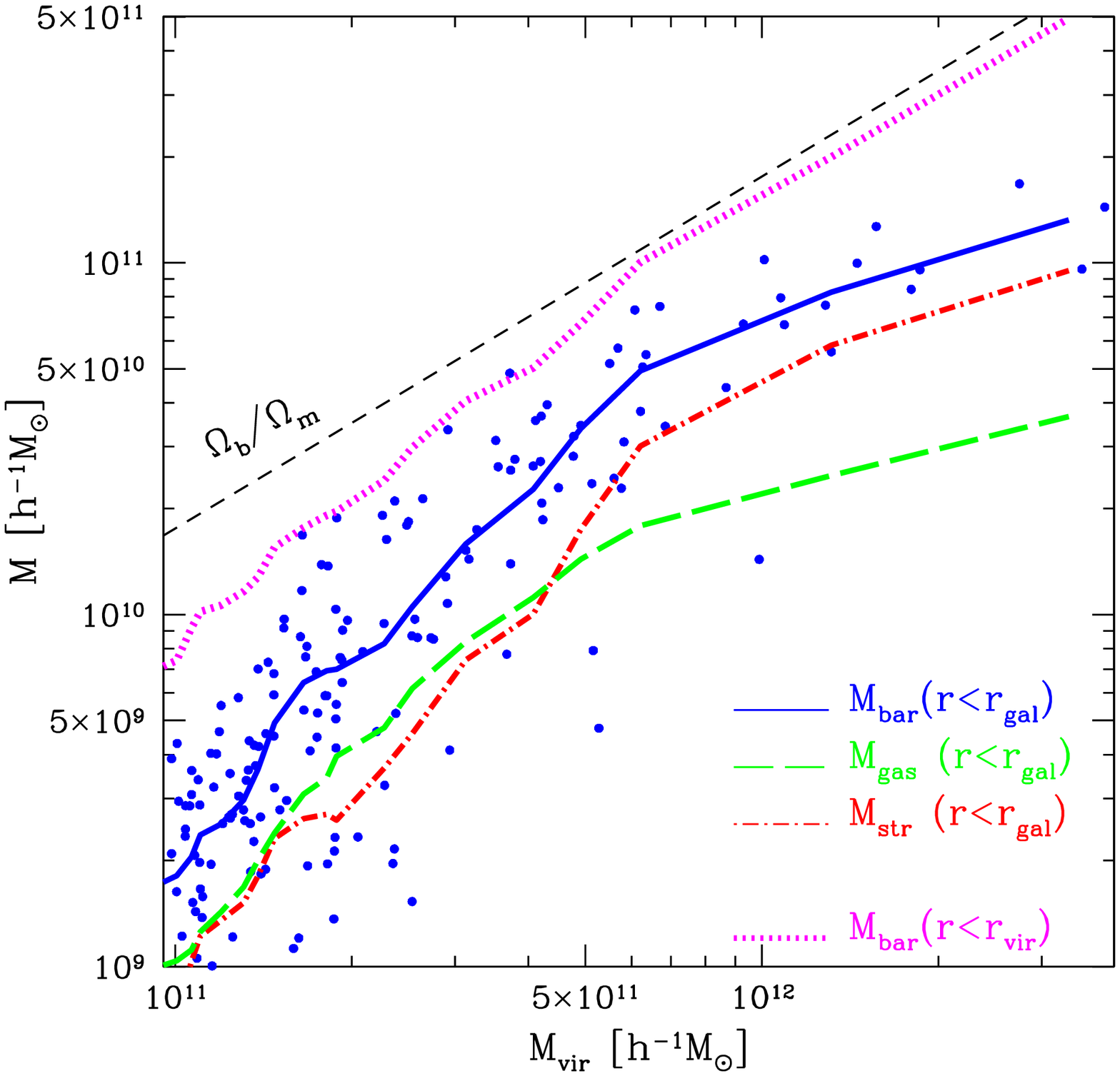}{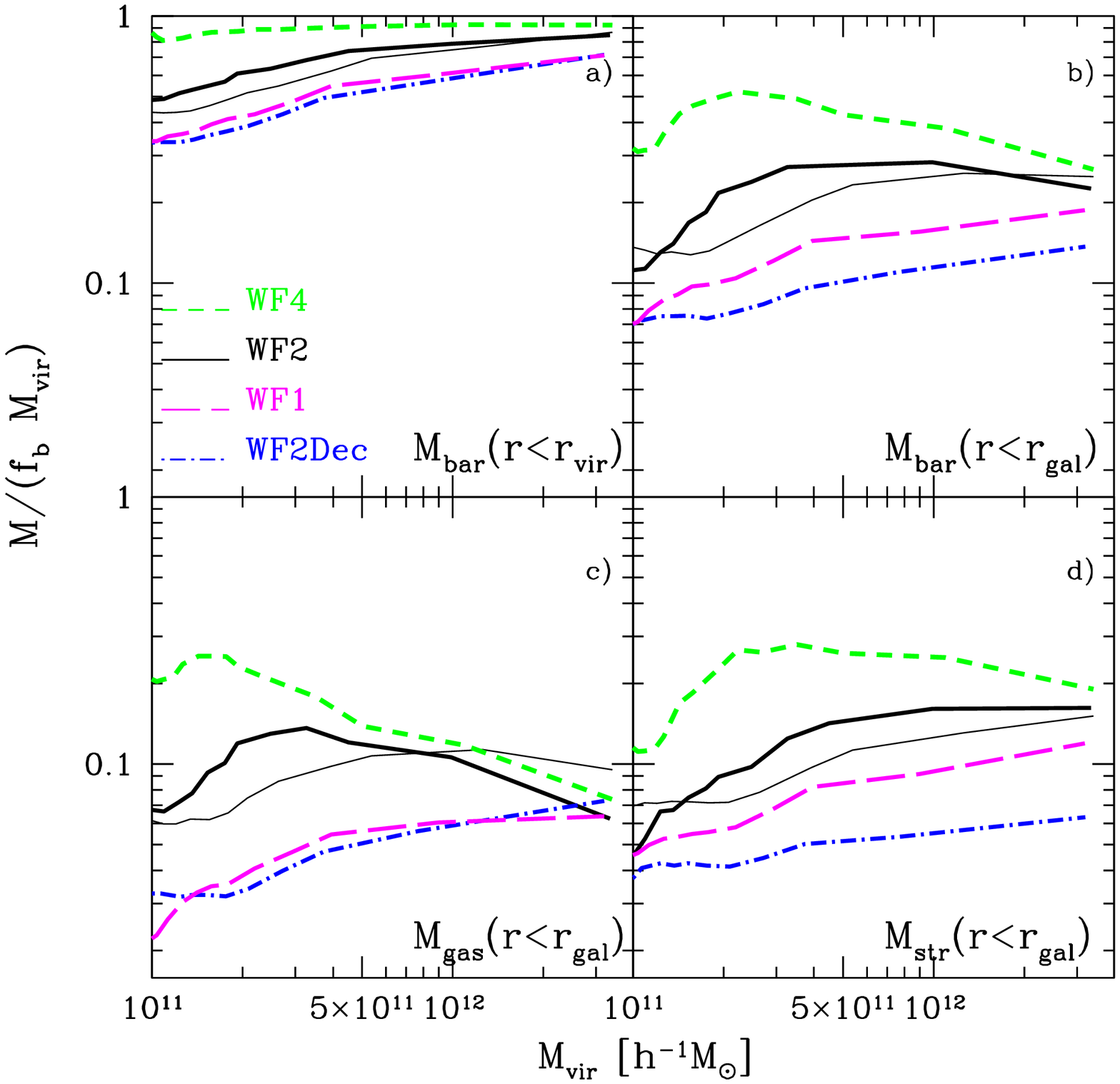}
\caption{{\it Left:} Galaxy mass as a function of halo virial mass for
  galaxies in the WF2 run. The various curves are described in the
  text, and trace medians in equal-number bins of virial
  mass. Dot-dashed, dashed, and solid lines correspond to stars, gas
  and all baryons (=gas+stars), respectively. Individual values are
  shown for the gas+stars case, in order to illustrate the
  halo-to-halo scatter. {\it Right:} Galaxy mass in units of the total
  baryon mass of the halo. Different curves in each panel correspond
  to runs with different feedback implementation.  Clockwise from the top
  left, each panel shows (a) the baryonic mass within the virial
  radius; (b) the baryonic mass in the central galaxy (i.e., within
  $r_{\rm gal}$); (c) the stellar mass within $r_{\rm gal}$, and (d)
  the gaseous mass within $r_{\rm gal}$.  The thin solid line
  correspond to a run with the same parameters as WF2 but with 8 times
  fewer particles, a good indicator of numerical convergence.}
\label{fig:fig1}
\end{figure}

For the range of halo masses investigated here the effects of feedback
(in terms of how effectively it regulates star formation and/or
removes gas from galaxies) increase with $v_{\rm w}$ for fixed $\eta \times
v_{\rm w}^2$. We refer to each of the 4 runs, in order of increasing overall
feedback efficiency, as: WF4 ($\eta=4$ particles and $v_{\rm w}=424$ km/s), WF2
($\eta=2$ and $v_{\rm w}=600$ km/s), WF1 ($\eta=1$ and $v_{\rm w}=848$ km/s) and
WF2Dec. The latter is equivalent to WF2 but ``wind'' particles are
temporarily decoupled from the hydrodynamical equations (see Dalla
Vecchia \& Schaye 2008 for details).

Our simulated galaxy sample consists of all galaxies assembled (at
$z=2$) at the centers of dark halos in the mass range $\rm M_{\rm
  vir}=10^{11}$ to $3 \times 10^{12} \, h^{-1} \rm M_\odot$. These
halos contain between $50,000$ and $500,000$ particles within the
virial radius (defined as the radius where the mean inner density is
$\Delta=178$ times the critical density of the universe at $z=2$). The
gravitational softening is never greater than $0.5 \, h^{-1} \rm kpc$
(physical). With these definitions, the simulated galaxy sample in
each run contains roughly $\sim 170$ objects.

\section{Results}
\subsection{Mass and angular momentum content}

The left panel of Fig.~\ref{fig:fig1} shows, as a function of halo
virial mass, the baryonic mass in various components for galaxies
identified in the WF2 run. The (top) dashed straight line indicates
the baryon mass associated with each halo through the universal baryon
fraction $f_b=\Omega_b/\Omega_{\rm M}$. The bottom four curves show
the baryon mass in different regions of the halo. The dotted line
corresponds to all baryons within $r_{\rm vir}$. This shows that halos
at the high-mass end of the considered range are able to retain all
baryons but that those at the low mass end have experienced a net loss
of about $\sim 50\%$ of their baryons. This illustrates the decreasing
efficiency of feedback-driven winds in the deeper potential wells of
more massive halos.

The effect is even more pronounced when considering the total amount
of baryons locked in the central galaxy (i.e., those within $r_{\rm
  gal} = 0.1 \, r_{\rm vir}$ from the halo centre). The dots in the
left panel of Fig.~\ref{fig:fig1} show that the total baryon mass of
the central galaxy is not simply proportional to the halo mass. The
relation has substantial scatter and it varies systematically as a
function of $M_{\rm vir}$. (All curves in this plot trace the median
in equal-number bins of virial mass.)  At low $M_{\rm vir}$ central
galaxies collect on average only about $10\%$ of the baryon mass
associated with the halo through the universal baryon fraction.  This
fraction rises to almost $50\%$ for $6\times 10^{11} h^{-1} M_\odot$
halos, but appears to decline in more massive halos.

The decline at the high-mass end is likely a result of the longer
cooling times characteristic of massive halos, which hinder the
assembly of massive central galaxies. The bottom two curves indicate
how the baryon mass of central galaxies splits into gas (dashed) and
stars (dot-dashed). Gas and stars contribute about equally at low
masses, but central galaxies in high-mass halos are predominantly
stellar. This is presumably a result of the inability of feedback to
delay or prevent the efficient conversion of gas into stars in halos
with a deep potential well. The effect on these trends of different
feedback prescriptions may be seen on the right panel of
Figure~\ref{fig:fig1}. Clearly, variations in feedback implementation have a
large effect on the mass (and gas fraction) of central galaxies. In
low mass halos (e.g., $M_{\rm vir} \sim 2\times 10^{11} h^{-1}
M_\odot$) WF4 and WF2Dec galaxies differ by more than a factor of 5,
illustrating the wide range of feedback efficiency surveyed in the OWLS
simulations.

\begin{figure}[!ht]
\plottwo{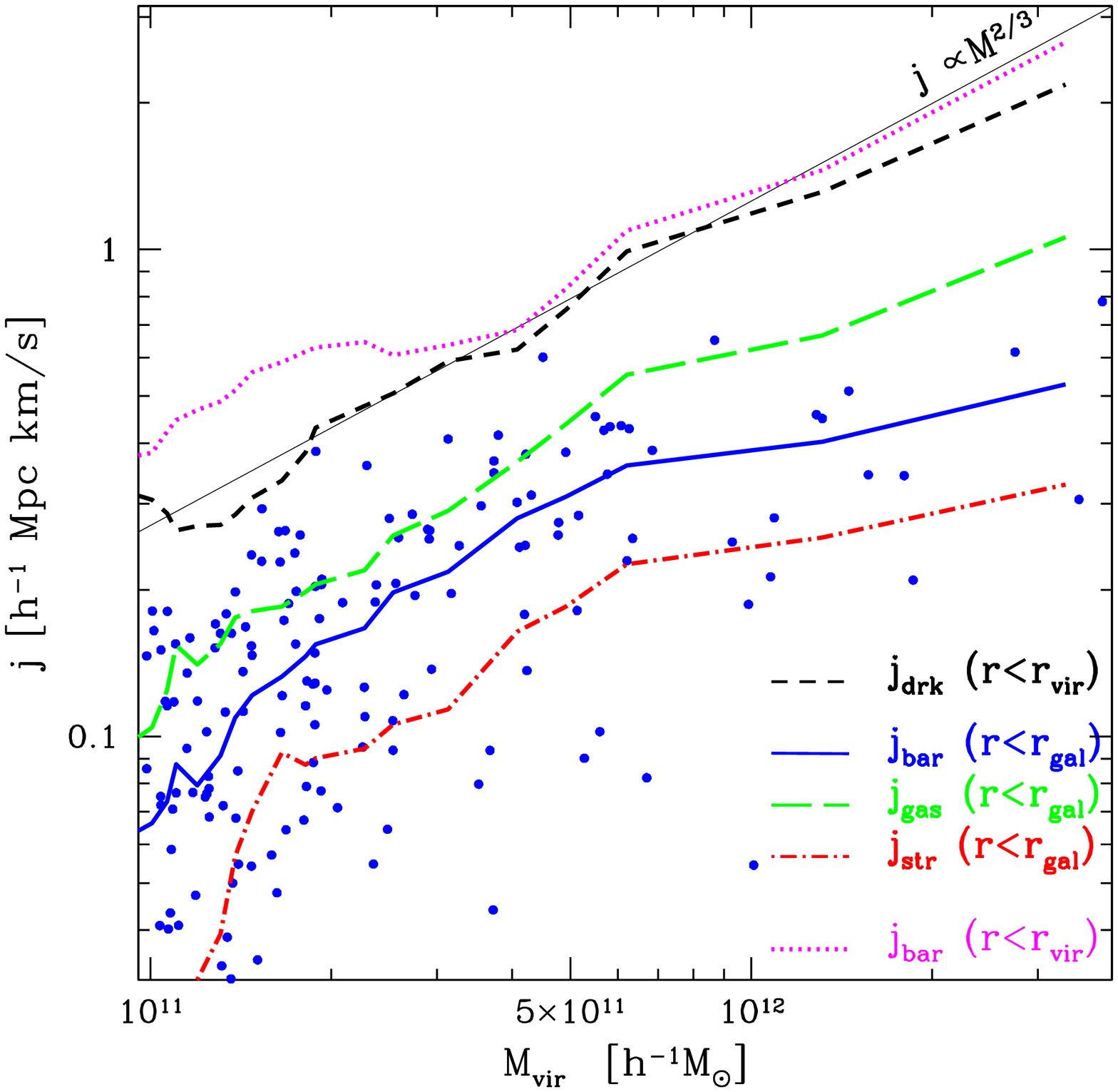}{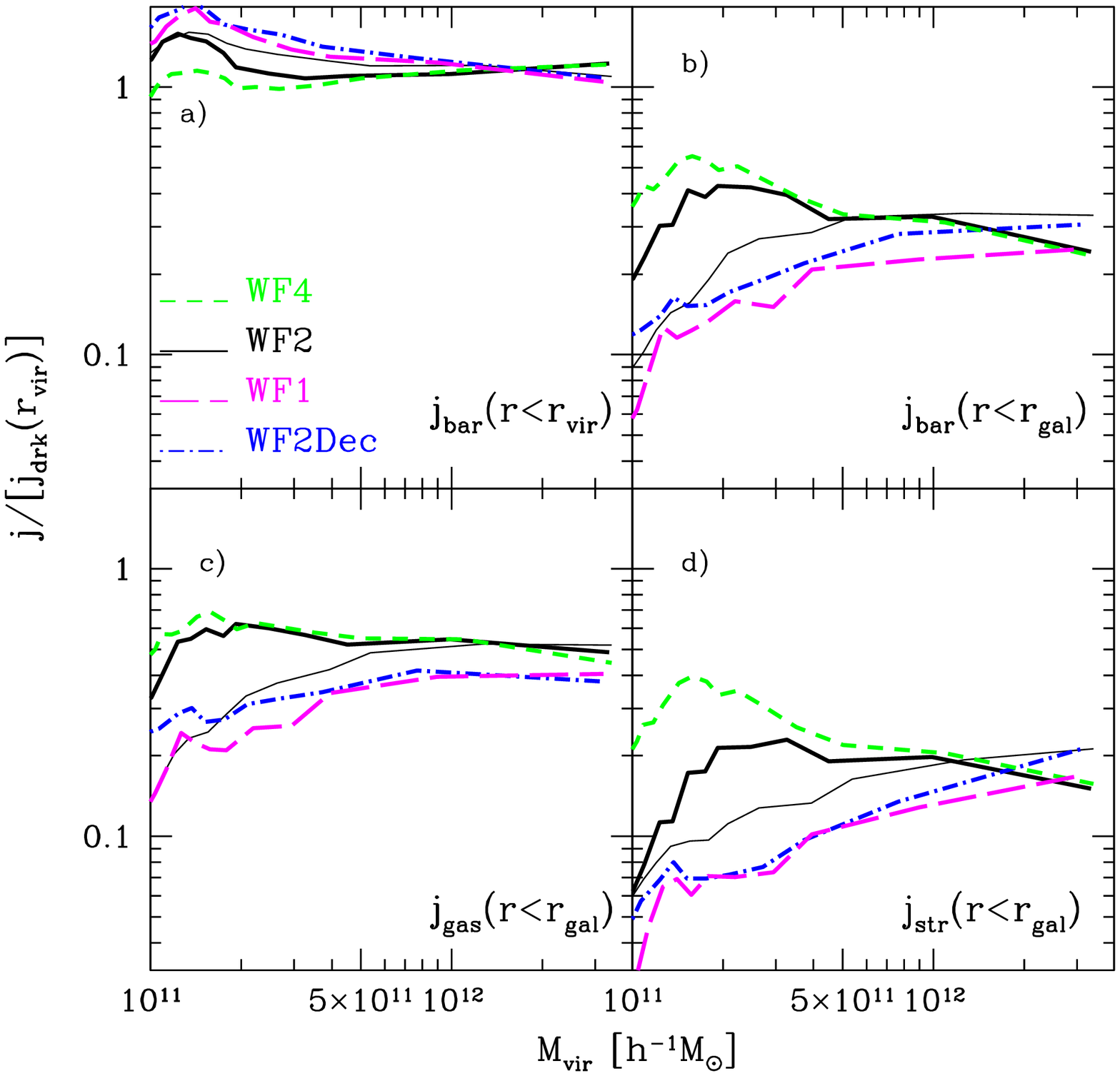}
\caption{{\it Left:} Specific angular momentum ($j$) as a function of
  halo virial mass in the WF2 galaxy sample. Line types are as in
  Fig.~\ref{fig:fig1}. The dashed black curve shows that the specific
  angular momentum of the dark halo scales roughly as $M^{2/3}$, as
  expected for systems with constant dimensionless spin parameter
  $\lambda$. The bottom curves show the specific angular momentum of
  stars (red dot-dashed), gas (green long-dashed) and all baryons
  (blue solid) within $r_{\rm gal}$. The mean angular momentum of all
  baryons within the virial radius ($j_{\rm bar}(r<r_{\rm vir})$) is
  shown in magenta dotted line. {\it Right:} Specific angular momentum
  of various components expressed in units of the halo's, as a
  function of halo virial mass. Different lines in each panel
  correspond to various feedback prescriptions, as described in the
  caption to Fig.~\ref{fig:fig1}.}
\label{fig:fig2}
\end{figure}

The left panel of Fig.~\ref{fig:fig2} shows the {\it specific} angular
momentum as a function of halo mass for various components in the WF2
galaxy sample. The dark matter angular momentum scales roughly as
$j_{\rm drk}\propto M^{2/3}$; the expected scaling for systems with
similar dimensionless spin parameter $\lambda$. This scaling is
followed closely by all baryons within the virial radius (dotted
line), a result that illustrates the fact that both baryons and dark
matter were torqued and spun by the same amount during the early
expansion phase of each system (see, e.g., Navarro et al 2004).

By comparison, the baryons in the central galaxies have much lower
specific angular momentum; a factor of $\sim 3$ lower on average (solid
curve and dots in the left panel of Fig.~\ref{fig:fig2}). It is also
clear from this figure that the gas (dashed curve) has typically
higher angular momentum than the stars (dot-dashed). This reflects the
higher efficiency of transformation of gas into stars in the central
(denser) regions of a galaxy and implies that gaseous disks will be,
on average, larger than stellar ones. As we discuss below, it is
important to take this effect into account when comparing theoretical
predictions with observations.

These trends are common to all our feedback implementations, as may be
verified by inspecting the right panel of Fig.~\ref{fig:fig2}. Note,
however, that feedback affects the mass and angular momentum of
central galaxies in different ways.  Indeed, compare the central
galaxy masses (top-right panel of Fig.~\ref{fig:fig1}) and spins (top
right panel of Fig.~\ref{fig:fig2}) of models WF4 and WF2Dec, our most
and least efficient feedback runs, respectively. In high-mass halos
the baryonic mass of WF4 central galaxies is roughly $\sim 2$-$3$
larger than in WF2Dec, but its specific angular momentum remains
largely unchanged.

\begin{figure}[!ht] 
\plotone{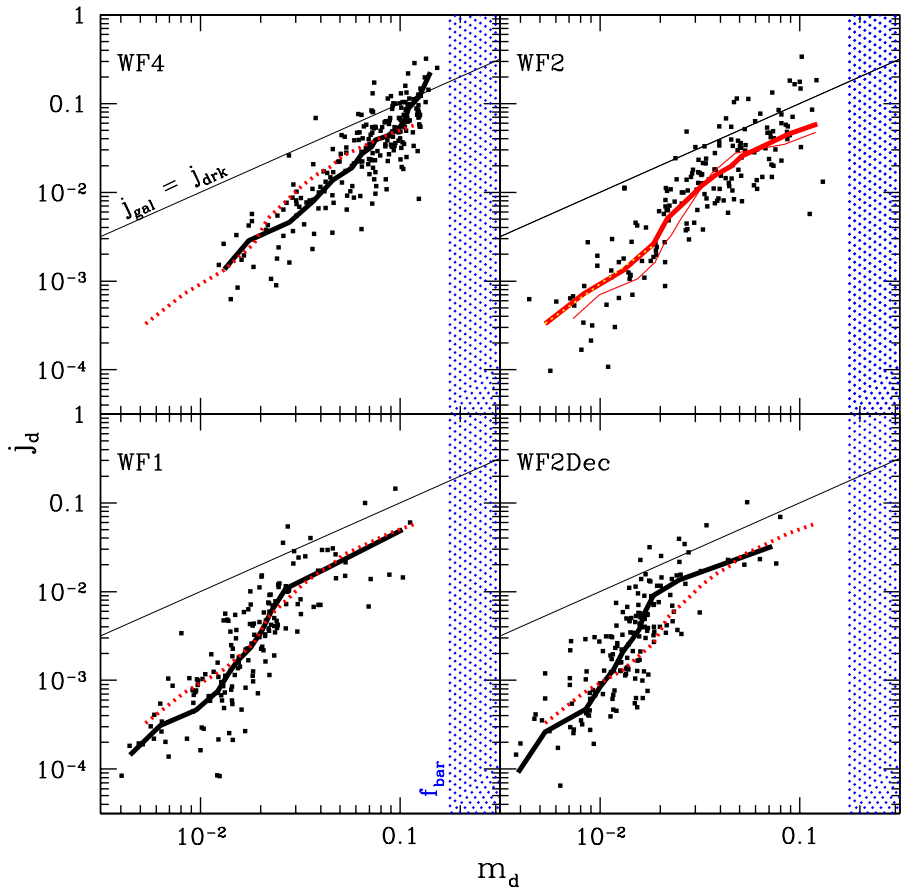} 
\caption{Baryonic mass $m_d=M_{\rm gal}/M_{\rm vir}$ and angular
  momentum $j_d=J_{\rm gal}/J_{\rm vir}$ of central galaxies, in units
  of the halo virial mass and angular momentum. Each panel corresponds
  to a different feedback prescription.  Thick lines in each panel
  show the median $j_d$ in equal-number $m_d$ bins for each run. (The
  thin line in the WF2 panel corresponds to a similar simulation with
  8 times fewer particles.)  The red dotted line, corresponding to
  WF2, is repeated in all panels, and highlights our result that the
  $m_d$-$j_d$ relation is approximately independent of feedback
  efficiency. The blue shaded area shows the region where the central
  galaxy mass would exceed the baryon mass within $r_{\rm vir}$; i.e.,
  $m_d>f_{\rm bar}$. No galaxies are expected (or seen) in this region
  of the plot.}
\label{fig:fig3}
\end{figure}

This non-linear interplay between the mass and spin of the baryons
that are assembled into central galaxies is explored in
Fig.~\ref{fig:fig3}, where we show the galaxy's mass ($m_d=\rm M_{\rm
  gal}/\rm M_{\rm vir}$) and angular momentum ($j_d=J_{\rm gal}/J_{\rm
  vir}$), each expressed in units of the corresponding halo
values. Semianalytic models of galaxy formation typically assume that
the specific angular momentum of the central galaxy is similar to that
of its host halo; i.e., that $m_d=j_d$. In the MMW98 formalism, these
parameters enable prediction of the exponential scalelength of a disk
galaxy once the virial radius ($r_{\rm vir}$) and spin parameter
($\lambda$) of a halo are specified,
\begin{equation}
R_d={(2f_c)}^{-1/2}(j_d/m_d)\, \lambda \,
r_{\rm vir} \, f_R(j_d,m_d,c,\lambda),
\end{equation}
where $f_R$ and $f_c$ are factors that account for the
concentration and ``contraction'' of the dark halo.

\begin{figure}[!ht]
\plottwo{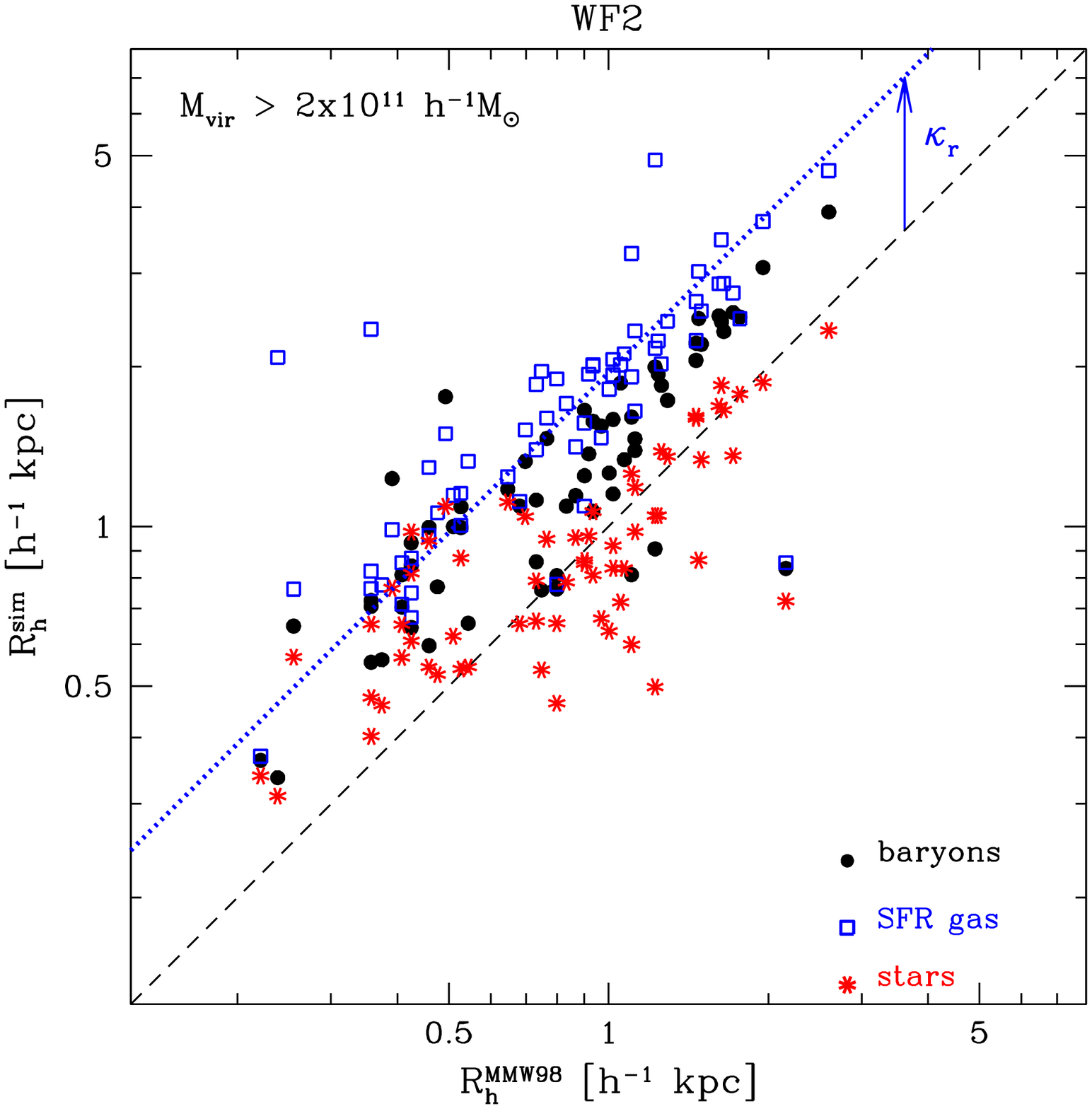}{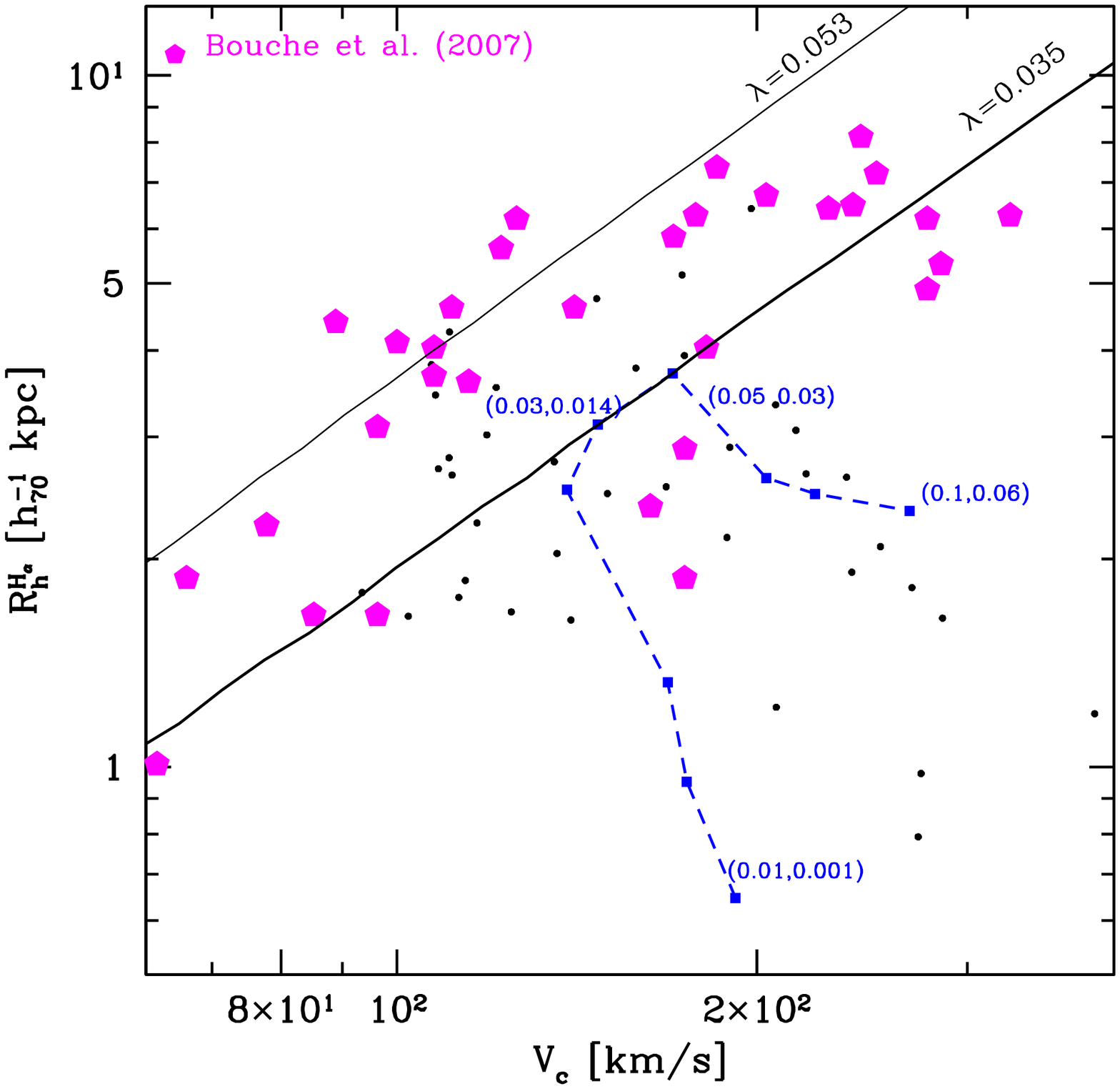}
\caption{{\it Left:} Cylindrical half-mass radii predicted by the
  MMW98 formalism (x-axis) versus projected half-mass radius of
  simulated galaxies. Stars are shown by red asterisks, star-forming
  gas by empty blue squares, and all baryons by solid black dots. {\it
    Right:} The size-velocity plane of $z=2$ SINS galaxies (magenta
  pentagons), together with the prediction of MMW98 for disks in a halo
  with $V_{\rm vir}=150$ km/s, average concentration, and spin
  parameter $\lambda=0.035$. The disk parameters $m_d$ and $j_d$ are
  assumed to follow the median $m_d$-$j_d$ relation found in the
  simulations. The solid black curves outline the size-velocity
  relation expected for disks with $(m_d,j_d)=(0.05,0.03)$ in halos
  with $\lambda=0.035$ and $0.053$, respectively.  Black dots
  correspond to a subset of our simulated galaxies with $m_d$-$j_d$
  values close to the values that maximize disk size: $m_d=0.05$ and
  $j_d=0.03$. These show that extended disks are indeed present at
  $z=2$ in our simulations.}
\label{fig:fig4}
\end{figure}

The $m_d$-$j_d$ relation in our simulations is shown in
Fig.~\ref{fig:fig3}, where each panel corresponds to a different
feedback implementation. A few things are worth noting in this
figure. (i) Simulated galaxies deviate strongly from the $m_d=j_d$
line, which assumes that galaxies and halos have the same specific
angular momentum (i.e., $j_{\rm gal}=j_{\rm drk}$).  (ii) The scatter
in the relation is large, and in every panel there are cases where
$j_{\rm gal}>j_{\rm drk}$. These are galaxies where feedback has been
able to push out a substantial fraction of low-angular momentum
baryons. As a result, those that remain attached to the central galaxy
have specific angular momentum higher than the system as a
whole. (iii) Feedback efficiency largely determines $m_d$: for
example, WF4 galaxy masses are on average 3 to 5 times larger than
WF2Dec galaxies. This may be seen as a horizontal shift of points
between the corresponding panels. (iv) The $m_d$-$j_d$ relation is
approximately independent of feedback. This is illustrated by the
dotted curve, which is repeated in each panel, and which reproduces
the median $j_d$ as a function of $m_d$ in the WF2 run. This curve
approximates reasonably well the $j_d$-$m_d$ relation in {\it all}
panels. In other words, for the feedback implementation adopted in our
simulations the fraction of baryons that collects into the central
galaxy is the primary factor determining the angular momentum content
this galaxy retains.

\subsection{Implications for disk galaxy sizes}

We may also use our simulations to check the validity of the MMW98
formula for predicting galaxy sizes. This is shown in the left
panel of Fig.~\ref{fig:fig4}, where we plot the predicted
(cylindrical) half-mass radius ($R_h^{\rm MMW98}$) versus the
projected half-mass radius of stars (red asterisks), star-forming gas
(empty blue squares) and baryons (black circles), for galaxies
identified in the WF2 simulation.

The agreement is quite good for the stellar component, which is 
striking given the number of simplifying assumptions that go into the
prediction and that are not matched in detail by the simulations. For
example, the MMW98 prediction assumes that all baryons are in a
centrifugally-supported exponential disk; simulated galaxies, on the
other hand, exhibit a very wide range of disk-to-spheroid ratio.
Interestingly, when scaled by a constant factor $\kappa_r \sim 2$ the
MMW98 formula also predicts rather accurately the size of the
star-forming {\it gaseous} disks (see upper dotted curve in the left
panel of Fig.~\ref{fig:fig4}).

We can combine this result with the $m_d$-$j_d$ relation discussed
above to compute the expected size of (gaseous) disks at $z=2$.  As
shown in the right panel of Fig.~\ref{fig:fig4}, the disk size and
rotation speed depend sensitively on $j_d$ and $m_d$. The jagged curve
in this panel shows how the size of gaseous disks formed in a halo of
{\it fixed} virial velocity $v_{\rm vir}=150$ km/s changes as $j_d$
and $m_d$ are varied. The halo is assumed to have average
concentration for its mass, and average spin, $\lambda=0.035$. The
jagged curve maps, in the disk size-rotation speed plane, variations
in ($m_d$,$j_d$) chosen to lie along the relation shown by the dotted
line in Fig.~\ref{fig:fig3}, from ($0.01$,$0.001$) to ($0.1$,$0.06$).
The disk radius depends critically on the value of $j_d$ and $m_d$:
{\it disk sizes vary by a factor of $\sim 8$, even though the halo
  mass, concentration, and spin are fixed.} 

For small values of $j_d$ the central disk has very little angular
momentum, and therefore a rather small size. As $j_d$ (and $m_d$)
increases, the disk grows steadily in size. Disk radii are maximized
for $m_d=0.05$ and $j_d=0.03$ in this example. Increasing $j_d$ further
results in massive disks (because of the concurrent rise in $m_d$)
with very high rotation speeds. Since angular momentum scales as disk
size times rotation speed, these massive disks end up being actually
smaller, despite their larger $j_d$. Clearly, predicting the size of
disks is a rather uncertain endeavour unless one is able to constrain
the values of the parameters $j_d$ and $m_d$ to within a very narrow
range.

As may be seen in Fig.~\ref{fig:fig4}, disks in $V_{\rm vir}=150$ km/s
halos may reach a size of order $\sim 4 \, h_{70}^{-1}$ kpc,
comparable to the size of disks in the SINS sample. (SINS data are
shown by large pentagons in the right-hand panel of
Fig.~\ref{fig:fig4}.) Larger disks may also exist, for example, in
more massive (intrinsically larger) halos, or in halos with
higher-than-average $\lambda$, or in systems where galaxies have
retained an unusually high fraction of its angular momentum (i.e.,
those that scatter {\it above} the mean $m_d$-$j_d$ relation in
Fig.~\ref{fig:fig3}). Taking, as an example, the WF2 run, we find, at
$z=2$, about $5\times 10^{-3}$ gaseous disks exceeding $\sim 2 \,
h_{70}^{-1}$ kpc (physical) in size per cubic $h^{-1}$ Mpc
(comoving). These numbers are in rough agreement with the findings of
the SINS collaboration (Bouch{\'e} et al 2007). We hasten to add that
these are not ``typical'' galaxies at that redshift, but rather some
of the largest ones present then. The selection of SINS galaxies
favors large disks in order to enhance the probability of obtaining a
resolved velocity map, and therefore they cannot be considered
``average'' at that redshift either. The presence of a (reasonable!)
number of extended disks at $z=2$ might thus not require unusually
high halo spins nor present an insurmountable challenge to the LCDM
structure formation paradigm.




\end{document}